\documentclass[aps,prd,twocolumn,superscriptaddress,10pt,noshowpacs]{revtex4}  
\usepackage{hyperref}
\hypersetup{plainpages=false,colorlinks=true,citecolor=blue,linkcolor=blue,urlcolor=blue,filecolor=green,bookmarksopen=true}
\pdfoutput=1
\usepackage[active]{srcltx}
\usepackage{graphicx}
\usepackage[utf8]{inputenc}
\usepackage{amsmath}
\usepackage{times,txfonts}
\usepackage{float}

\usepackage{color}

\begin{document}

\title{Lorentz-violating extension of scalar QED at finite temperature}

\author{M. C. Ara\'{u}jo}
\email{michelangelo@fisica.ufc.br}
\affiliation{Universidade Federal do Cear\'a (UFC), Departamento de F\'isica,\\ Campus do Pici, Fortaleza - CE, C.P. 6030, 60455-760 - Brazil.}
\author{J. Furtado}
\email{job.furtado@ufca.edu.br}
\affiliation{Universidade Federal do Cariri (UFCA), Av. Tenente Raimundo Rocha, \\ Cidade Universit\'{a}ria, Juazeiro do Norte, Cear\'{a}, CEP 63048-080, Brasil}
\author{R. V. Maluf}
\email{r.v.maluf@fisica.ufc.br}
\affiliation{Universidade Federal do Cear\'a (UFC), Departamento de F\'isica,\\ Campus do Pici, Fortaleza - CE, C.P. 6030, 60455-760 - Brazil.}
\affiliation{Departamento de F\'{i}sica Te\'{o}rica and IFIC, Centro Mixto Universidad de Valencia - CSIC. Universidad de Valencia, Burjassot-46100, Valencia, Spain.}

\date{\today}

\begin{abstract}
In this work, we calculate the one-loop self-energy corrections to the gauge field in scalar electrodynamics modified by Lorentz-violating terms within the framework of the standard model extension (SME). We focus on both $CPT$-even and $CPT$-odd contributions. The kinetic part of the scalar sector contains a $CPT$-even symmetric Lorentz-breaking tensor, and the interaction terms include a vector contracted with the usual covariant derivative in a gauge-invariant manner. We computed the one-loop radiative corrections using dimensional regularization for both the $CPT$-even and $CPT$-odd cases. Additionally, we employed the Matsubara formalism to account for finite temperature effects.
\end{abstract}

\maketitle

\section{Introduction}

When we try to use Quantum Theory and General Relativity concepts to investigate a physical system confined to length scales in the order of Planck length, we can produce an energy density large enough to affect the spacetime structure itself \cite{Garay:1994en}. One of the possible effects of quantum gravity that can leave fingerprints on scales of much greater length than the Planck length is the violation of Lorentz symmetry \cite{Kostelecky:1988zi,Kostelecky:1991ak,Bojowald:2004bb,Carroll:2001ws}. A theoretical framework capable of exploring the Lorentz-violating effects, allowing the calculation of quantities measured in experiments, is the Standard-Model Extension (SME) \cite{Colladay:1996iz,Colladay:1998fq}.

Several properties of $CPT$ and/or Lorentz violation (LV) theories based on SME have been intensively examined in the Literature over the last years. One can cite, for example, spinor solutions of the Dirac equation for atomic systems \cite{Bluhm:1998rk, Phillips:2000dr,Ferreira:2007gnn}, tree-level propagator \cite{Casana:2018rhg,Maluf:2018jwc,Ferreira:2019lpu}, meson scattering at finite temperature \cite{Araujo:2022qke}, analyzing stability, microcausality, and unitarity \cite{Kostelecky:2000mm}, radiative corrections \cite{Jackiw:1999yp,Coleman:1998ti,Chung:1999gg,PerezVictoria:1999uh,Chung:2001mb, Furtado:2014cja}, modified dispersion relations \cite{Reis:2016hzu}, polarization vectors, and vacuum birefringence \cite{Klinkhamer:2008ky,Kostelecky:2009zp}, among others (see, e.g., \cite{Kostelecky:2008bfz}, and references therein).

One of the most interesting subjects of these works is the radiative generation of LV terms belonging to the SME. This issue was studied at the end of the '90s in Ref. \cite{Jackiw:1999yp}. Also, there was already some discussion of this same issue in Ref. \cite{Colladay:1998fq}, as well as by Coleman and Glashow in Ref. \cite{Coleman:1998ti}. In those works, it is shown that the CPT-odd or Carroll–Field–Jackiw (CFJ) term $\epsilon^{\mu\nu\alpha\beta}(k_{AF})_{\mu}A_{\nu}F_{\alpha\beta}$, present in the gauge sector of the SME, is induced by radiative corrections from a fermion LV term like $b_{\mu}\bar{\psi}\gamma^{\mu}\gamma_{5}\psi$. The one-loop self-energy correction obtained has the form $\Pi^{\mu\nu}=\lambda \epsilon^{\mu\nu\alpha\beta}b_{\alpha}p_{\beta}$, whose coefficient $\lambda$ is not unambiguously determined; it explicitly depends on the regularization prescription adopted to make the Feynman integrals convergent \cite{Jackiw:1999qq}. The radiative induction of the CFJ term was also studied in the context of finite temperature \cite{Mariz:2005jh,Nascimento:2007rb}, and it was also verified that the induced CFJ term is non-analytical at finite temperature \cite{Assuncao:2016fko}. In fact, the non-analyticity of the Chern-Simons term is not a particular feature of four-dimensional theories since it is also exhibited in five dimensions \cite{Assuncao:2020cdo}. The Chern-Simons modified gravity was also induced in the finite temperature regime in Ref. \cite{Assuncao:2018jkq}, and higher-derivatives Lorentz-breaking terms were discussed in Refs. \cite{Celeste:2016tar, Leite:2013pca}.

The radiative generation of other Lorentz-violating terms has been demonstrated extensively in the last few years. In Refs. \cite{Gomes:2009ch,Scarpelli:2013eya}, it was shown that the aether-like terms \cite{Carroll:2008pk} can be induced by convenient couplings to the spinor fields. Higher-dimensional gauge terms, like the Myers-Pospelov term, were obtained in Refs. \cite{Mariz:2010fm,Mariz:2011ed}. In Ref.  \cite{Casana:2013nfx}, it was shown the radiative generation of the full $CPT$-even term of the SME electrodynamics, $(K_{F})_{\mu\nu\alpha\beta}F^{\mu\nu}F^{\alpha\beta}$, through a nonminimum coupling between the fermionic and gauge fields. Also, in the mentioned work, some upper bounds on the magnitude of the $CPT$-even for the nonbirefringent and birefringent coefficients were presented to the tensor $(K_{F})$ coefficients.

In this work, we calculate the one-loop self-energy corrections to the gauge field in scalar electrodynamics modified by Lorentz-violating terms in the framework of the standard model extension (SME). We focus our attention on both $CPT$-even and $CPT$-odd contributions. The kinetic part of the scalar fields contains a $CPT$-even symmetric Lorentz-breaking tensor, and the interaction terms include a vector contracted with the usual covariant derivative in a gauge-invariant mode. For both $CPT$-even and $CPT$-odd cases, the one-loop radiative corrections were computed using dimensional regularization, and we employed the Matsubara formalism in order to separate the zero and finite temperature contributions. 

This paper is organized as follows: In the next section, we obtained the general expression for the one-loop effective action considering both Lorentz-violating contributions, namely, the $CPT$-even and the $CPT$-odd. In two separate subsections, we individually computed the zero and finite temperature contributions from the $CPT$-even and the $CPT$-odd terms. In section III, we present our final remarks. We have considered $g^{\mu\nu}=diag(1,-1,-1,-1)$ during the calculations.

\section{One-loop photon self-energy tensor}

The model we are considering consists of a minimal coupling between the scalar fields of the Lorentz-violating extension of the standard model, recently proposed by Kostelecky and Edwards \cite{Edwards:2018lsn}, with the electromagnetic field including the usual Maxwell term. Hence the Lagrangian density describing the system is
\begin{eqnarray}\label{lphipluslg}
\mathcal{L}=\mathcal{L}_{\phi}+\mathcal{L}_G ,\label{Lag}
\end{eqnarray}where,
\begin{eqnarray}
\!\!\!\!\!\mathcal{L}_{\phi}\!\!&=&\!\!G^{\mu\nu}(D_{\mu}\phi)^{\dagger}D_{\nu}\phi-m^2\phi^{\dagger}\phi-\frac{i}{2}[\phi^{\dagger}\hat{k}_a^{\mu}D_{\mu}\phi-\phi\hat{k}_a^{\mu}(D_{\mu}\phi)^{\dagger}],\,\,\,\\
\!\!\!\!\!\mathcal{L}_G\!\!&=&\!\!-\frac{1}{4}F_{\mu\nu}F^{\mu\nu}.
\end{eqnarray}
The covariant derivative is defined as $D_{\mu}=\partial_{\mu}-ieA_{\mu}$, the tensor $G^{\mu\nu}=g^{\mu\nu}+(\hat{k}_c)^{\mu\nu}$ is composed by the Minkowski metric tensor $g^{\mu\nu}$ and a Lorentz-violating traceless constant tensor $(\hat{k}_c)^{\mu\nu}$. The tensor $(\hat{k}_c)^{\mu\nu}$ and the vector $\hat{k}_a^{\mu}$ promotes the violation of the Lorentz invariance by breaking the equivalence between particle and observer transformations. Such tensors, assumed to be constant, imply the independence of the space-time position, which yields translational invariance assuring the conservation of momentum and energy. 

Thus, the corresponding generating functional of the theory involves the functional integration over its degrees of freedom, as can be seen below,
\begin{eqnarray}
\!\!\!\!Z=\int DA_{\mu}D\phi^{\dagger} D \phi \,e^{i\int d^4x\mathcal{L}}=\int DA_{\mu}\,e^{-\frac{i}{4} \int d^4x F_{\mu\nu}F^{\mu\nu}+iS_{eff}},
\end{eqnarray}
where the one-loop effective action for the scalar fields can be written as,
\begin{eqnarray}\label{effectiveaction}
\nonumber S_{eff}&=&-i \mbox{Tr}\ln\left[G^{\mu\nu}p_{\mu}p_{\nu}-m^2 -\hat{k}_a^{\mu}p_{\mu}-e\hat{k}_a^{\mu}A_{\mu}\right.\\
&&\left.+eG^{\mu\nu}k_{\nu}A_{\mu}+2eG^{\mu\nu}A_{\mu}p_{\nu}+e^2G^{\mu\nu}A_{\mu}A_{\nu}\right].
\end{eqnarray}
Here, $\mbox{Tr}$ stands only for the trace over the space coordinate.

In order to single out quadratic terms on $A_{\mu}$, it is convenient to write the effective action in eq. (\ref{effectiveaction}) as
\begin{eqnarray}
S_{eff}=S_{eff}^{(0)}+\sum_{n=1}^{\infty}S_{eff}^{(n)},
\end{eqnarray}
where,
\begin{eqnarray}
S_{eff}^{(0)}&=&-i\mbox{Tr}\ln\left[p^2-m^2-\hat{k}_a^{\mu}p_{\mu}+\hat{k}_c^{\mu\nu}p_{\mu}p_{\nu}\right],\\
\nonumber S_{eff}^{(n)}&=&\frac{i}{n}\mbox{Tr}\left[G(p)\left(e\hat{k}_a^{\mu}A_{\mu}-eG^{\mu\nu}k_{\nu}A_{\mu}\right.\right.\\
&&\left.\left.-2eG^{\mu\nu}A_{\mu}p_{\nu}-e^2G^{\mu\nu}A_{\mu}A_{\nu}\right)\right]^n,
\end{eqnarray}
with $G(p)=(p^2-m^2-\hat{k}_a^{\mu}p_{\mu}+\hat{k}_c^{\mu\nu}p_{\mu}p_{\nu})^{-1}$. The contributions to the vacuum polarization come from the $n=1$ and $n=2$, i.e., second-order contributions in $e$. Hence, after the trace evaluation over the space coordinate we obtain
\begin{eqnarray}
    S_{eff}^{(1,2)}=-\frac{i}{2}\int\frac{d^4k}{(2\pi)^4}\Tilde{A}_{\mu}(-k)\Pi^{\mu\nu}(k)\Tilde{A}_{\nu}(k),
\end{eqnarray}
where 
\begin{eqnarray}
   \Pi^{\mu\nu}(k)=T^{\mu\nu}(k)+R^{\mu\nu}(k), 
\end{eqnarray} being $T^{\mu\nu}(k)$ and $R^{\mu \nu}(k)$ the polarization tensors due to the tensor and vector SME coefficients, respectively. At this point, it is convenient to treat each contribution separately.  
\subsection{$\hat{k}_c^{\mu\nu}$ contributions}

\begin{figure*}
    \centering
    \includegraphics[scale=0.8]{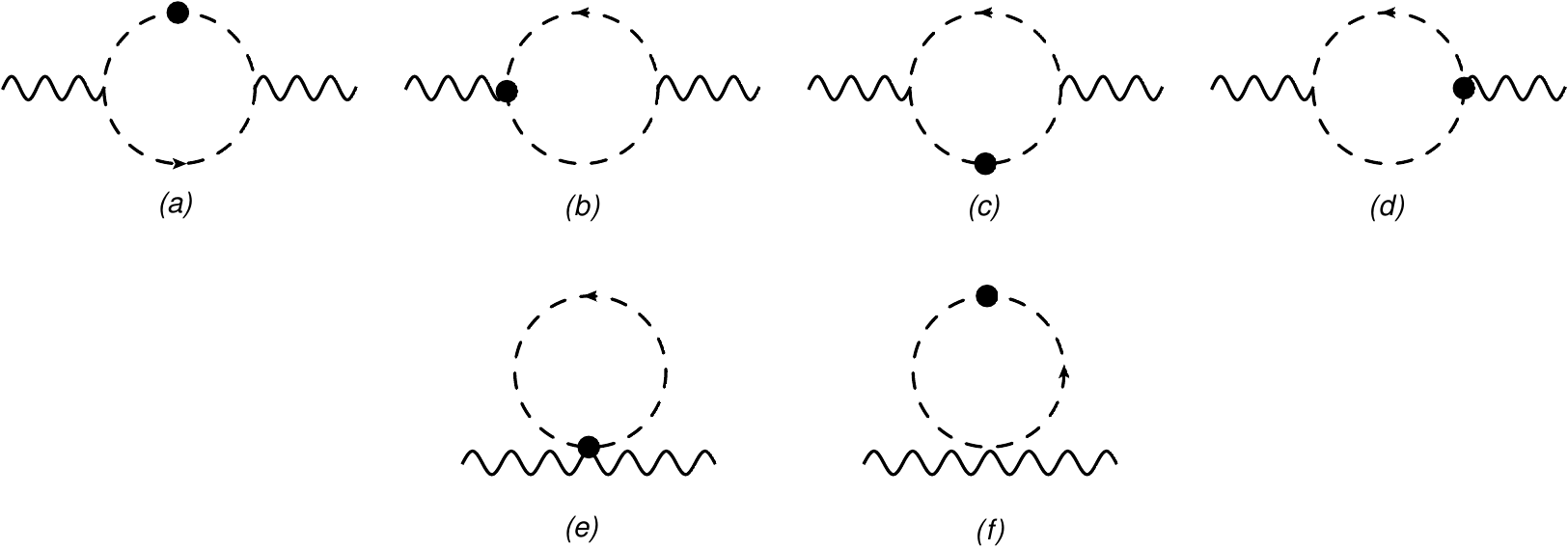}
    \caption{Feynman diagrams}
    \label{figkc}
\end{figure*}

In this section, we will calculate the vacuum polarization at finite temperature considering the coefficient $(\hat{k}_c)^{\mu\nu}$. The polarization tensor corresponding to the Feynman graphs depicted in Fig. \ref{figkc}  can be written as
\begin{equation}\label{t1t2t3t4prim}
    T^{\mu\nu}(k)=T_1^{\mu\nu}(k)+T_2^{\mu\nu}(k)+T_3^{\mu\nu}(k)+T_4^{\mu\nu}(k),
\end{equation} with
\begin{eqnarray}
  \label{eq1212eq}\!\!\!\!\!\!\!\!\!\!\!\!\!\!\!T_1^{\mu \nu}(k) &=& 2e^2(\hat{k}_c)^{\alpha \beta}\int \frac{d^4p}{(2\pi)^4} \frac{ p_{\alpha}p_{\beta}}{(q^2-m^2)(p^2-m^2)^2}\nonumber \\
  &\times& [g^{\mu \nu}(q^2-m^2)-(p^{\mu}+q^{\mu})(p^{\nu}+q^{\nu})], \\
\label{eq1313eq} \!\!\!\!\!\!\!\!\!\!\!\!\!\!\!T_2^{\mu \nu}(k) &=& e^2 (\hat{k}_c)^{\mu \alpha}\int \frac{d^4p}{(2\pi)^4} \frac{(p_{\alpha}+q_{\alpha})(p^{\nu}+q^{\nu})(p^2-m^2)}{(q^2-m^2)(p^2-m^2)^2}, \\
 \label{eq1414eq} \!\!\!\!\!\!\!\!\!\!\!\!\!\!\!T_3^{\mu \nu}(k) &=& e^2 (\hat{k}_c)^{\nu \alpha}\int \frac{d^4p}{(2\pi)^4} \frac{(p_{\alpha}+q_{\alpha})(p^{\mu}+q^{\mu})(p^2-m^2)}{(q^2-m^2)(p^2-m^2)^2},\\
\label{eq1515eq}  \!\!\!\!\!\!\!\!\!\!\!\!\!\!\!T_4^{\mu \nu}(k) &=& -e^2 (\hat{k}_c)^{\mu \nu}\int \frac{d^4p}{(2\pi)^4} \frac{(q^2-m^2)(p^2-m^2)}{(q^2-m^2)(p^2-m^2)^2},
\end{eqnarray} and $q=p+k$. Here, it is worth an observation: the diagrams $(a)$ and $(c)$ contribute with the same expression, and we are condensing into Eq. \eqref{eq1212eq} the expressions coming from the diagrams $(a)$, $(c)$ and $(f)$. Furthermore, Eqs. \eqref{eq1313eq}-\eqref{eq1515eq} come  from the diagrams $(b)$, $(d)$ and $(e)$, respectively. As we can see yet, we are considering only the first-order contribution in the coefficient $(\hat{k}_c)^{\mu\nu}$ since it is expected that the bounds on Lorentz violating coefficients to be small. 

In order to write the denominators in a more suitable way, we make use of Feynman parametrization and write
\begin{eqnarray}\label{tddimeni}
 \!\!\!\!\!\!\!\!\!\!\!\!\!\!\!T_i^{\mu \nu}(k) &=& 2e^2\mu^{4-D} \int_{0}^{1}dx \, 2(1-x)\int \frac{d^Dp}{(2\pi)^D} \frac{(N_i)^{\mu \nu}}{(p^2-M^2)^3},
\end{eqnarray} where $i=1,2,3,4$ and $M^2=m^2+x(x-1)k^2$ such that
\begin{eqnarray}
   \!\!\!\!\! (N_1)^{\mu \nu} &=& x^2(\hat{k}_c)^{\alpha \beta} k_{\alpha} k_{\beta} \lbrace g^{\mu \nu}[(1-x)^2k^2-m^2] - (1-2x)^2k^{\mu}k^{\nu} \rbrace \nonumber\\
     &+& \lbrace  4x(1-2x)[(\hat{k}_c)^{\mu \alpha}k_{\alpha}k^{\nu}+(\hat{k}_c)^{\nu \alpha}k_{\alpha}k^{\mu}] \nonumber\\ 
     &+& (\hat{k}_c)^{\alpha\beta}k_{\alpha}k_{\beta}g^{\mu \nu}(Dx^2-4x) \rbrace \frac{p^2}{D}-8(\hat{k}_c)^{\mu \nu}\frac{p^4}{D(D+2)}, \nonumber\\
    \!\!\!\!\! (N_2)^{\mu \nu} &=& \frac{(1-2x)^2}{2}(k^2x^2-m^2)(\hat{k}_c)^{\mu \alpha}k_{\alpha}k^{\nu}\nonumber \\
    &+& \lbrace (2x-1)[2x(D+4)-D](\hat{k}_c)^{\mu \alpha}k_{\alpha}k^{\nu}\nonumber \\
    &+& 4(x^2k^2-m^2)(\hat{k}_c)^{\mu \nu} \rbrace \frac{p^2}{2D}\nonumber \\
    &+& 24(\hat{k}_c)^{\mu \nu}\frac{p^4}{2D(D+2)},\\
    (N_3)^{\mu \nu} &=& (N_2)^{\nu \mu},\\
    \!\!\!\!\! (N_4)^{\mu \nu} &=& -\lbrace x^2(x-1)^2k^4-(2x^2-2x+1)m^2k^2+m^4\rbrace \nonumber\\
    &-& \lbrace k^2[2Dx(x-1)+D+4x(x-1)]-2Dm^2 \rbrace \frac{p^2}{D} \nonumber\\
    &-&p^4.
\end{eqnarray} Note that we have employed dimensional regularization to isolate the divergences. The procedure consists in extending the space-time from $4$ to $D$ dimensions, so that integration measure goes from $d^4p/(2\pi)^4$ to $\mu^{4-D}d^Dp/(2\pi)^D$, with $\mu$ being a mass regulator. Solving the momentum integrals and expanding the result around $\epsilon=4-D$ gives us the following result at zero temperature
\begin{eqnarray}
\nonumber T^{\mu\nu}(k)&=&-\frac{ie^2}{48\pi^2\epsilon}\left\{g^{\mu\nu}[2k_{\alpha}k_{\beta}(\hat{k}_c)^{\alpha\beta}-\kappa k^2]+2k^2(\hat{k}_c)^{\mu\nu}\right.\\
   &&+\left.\kappa k^{\mu}k^{\nu}-2k^{\nu}k_{\alpha}(\hat{k}_c)^{\alpha\mu}-2k^{\mu}k_{\alpha}(\hat{k}_c)^{\alpha\nu} \right\},
\end{eqnarray}
where $\kappa=g^{\alpha\beta}(\hat{k}_{c})_{\alpha\beta}$. The above expression is gauge invariant, as we can see from the contraction $k_{\mu}T^{\mu\nu}(k)=0$ as well as $T^{\mu\nu}(k)k_{\nu}=0$. Considering the fact that the tensor $(\hat{k}_c)^{\mu \nu}$ is traceless, the above expression reduces to

\begin{eqnarray} \label{tzerotempzero}
    \nonumber T^{\mu\nu}(k)&=&-\frac{ie^2}{24\pi^2\epsilon}\left\{k_{\alpha}k_{\beta}(\hat{k}_c)^{\alpha\beta}g^{\mu\nu}+k^2(\hat{k}_c)^{\mu\nu}\right.\\
   &&-\left.k^{\nu}k_{\alpha}(\hat{k}_c)^{\alpha\mu}-k^{\mu}k_{\alpha}(\hat{k}_c)^{\alpha\nu} \right\}.\label{T0}
\end{eqnarray}

\subsubsection{Finite temperature regime}

Now, to implement the finite temperature by the Matsubara formalism, we should first write Eq. \eqref{tddimeni} in Euclidean space by performing a Wick rotation, i.e., $p^0 \rightarrow ip^0_E$, $\, d^Dp \rightarrow id^Dp_E $, $\, p^2 \rightarrow -p^2_E = -[(p^0_E)^2+\mathbf{p}^2]$, and then make the changes $\int \frac{d^Dp}{(2\pi)^D} \rightarrow \frac{1}{\beta}\sum_{n=-\infty}^{\infty} \int \frac{d^dp}{(2\pi )^d} $ and $p^0_E \rightarrow \frac{2 n \pi }{\beta}$ due to periodic boundary conditions for bosons, with $d=D-1$.  From now on, the system is in thermal equilibrium with a temperature $T = \frac{1}{\beta}$. As a consequence, we have to solve integrals like
\begin{eqnarray}
    I_j = \frac{1}{\beta}\sum_{n=-\infty}^{\infty} \int \frac{d^dp}{(2\pi )^d} \frac{1}{(\mathbf{p}^2+\Delta)^{4-j}},
\end{eqnarray} where $\Delta = (\frac{2\pi}{\beta})^2(n^2+\xi^2)$, with $\xi = \frac{M \beta}{2\pi}$ and $j = 1, 2, 3$. Computing the integral over the momenta we obtain
\begin{eqnarray}\label{gammassoma}
   I_j&=& \frac{1}{(4\pi )^{\frac{d}{2}}}\frac{\Gamma \left( 4-j-\frac{d}{2} \right)}{\Gamma (4-j)}\frac{1}{\beta}\left(\frac{2\pi}{\beta}\right)^{d-8+2j}\nonumber \\
   &\times& \sum_{n=-\infty}^{\infty}\left( n^2+\xi^2 \right)^{-\left( 4-j-\frac{d}{2} \right)}. 
\end{eqnarray} Since the summation may present singularities in the limit $d\rightarrow 3$, we are going to use the Ford expression \cite{Ford:1979ds}, \begin{eqnarray}
    \sum_{n=-\infty}^{\infty} [(n+b)^2+a^2]^{-\lambda} = \frac{\sqrt{\pi}\Gamma\left( \lambda-\frac{1}{2} \right)}{\Gamma\left( \lambda \right)(a^2)^{\lambda-\frac{1}{2}}}+4\sin (\pi \lambda) f_{\lambda}(a,b),\nonumber\\
\end{eqnarray} where
\begin{eqnarray}
    f_{\lambda}(a,b) = \int_{|a|}^{\infty}\frac{dz}{(z^2-a^2)^{\lambda}}\,  Re \left[ \frac{1}{e^{2\pi (z+ib)}-1} \right], 
\end{eqnarray} is a valid function for $Re[\lambda]< 1$, aside from the poles at
$\lambda = \frac{1}{2}, -\frac{1}{2}, ...$,  in order to isolate the singularities. As a result, Eq. \eqref{gammassoma} takes the form
\begin{eqnarray}\label{ijmdbetaa}
 I_j&=& \frac{M^{d-7+2j}\Gamma\left( \frac{7-2j-d}{2} \right)}{(4\pi)^{\frac{d+1}{2}}\Gamma(4-j)}\nonumber\\
 &+& \frac{\Gamma\left( j-\frac{d}{2} \right)}{(4\pi)^{\frac{d}{2}}\Gamma(4-j)}\frac{1}{\beta}\left( \frac{2\pi}{\beta} \right)^{d-8+2j}\sin\left(\frac{\pi d}{2}\right)A_j(\xi,d),\nonumber\\
\end{eqnarray} where
\begin{eqnarray}
    A_1(\xi,d) &=& \frac{\pi^2}{\xi^2}F_3(\xi,d)-\frac{(d-2)(d-3)}{2\xi^2}F_2(\xi,d),\\
    A_2(\xi,d) &=& -2F_2(\xi,d),\\
    A_3(\xi,d) &=& \frac{8}{(d-4)(d-2)}F_1(\xi,d).
\end{eqnarray} Here, $F_1(\xi,d)$, $F_2(\xi,d)$, and $F_3(\xi,d)$ are temperature-dependent functions, in an arbitrary dimension $d$, given by
\begin{eqnarray}
    F_k(\xi, d) &=& \int_{|\xi|}^{\infty}\frac{dz}{(z^2-\xi^2)^{k-\frac{d}{2}}}[\coth(\pi z)-1],
\end{eqnarray} for $k=1,2$, and
\begin{eqnarray}
    F_3(\xi, d) &=& \int_{|\xi|}^{\infty}\frac{dz}{(z^2-\xi^2)^{1-\frac{d}{2}}} \coth (\pi  z) \, \text{csch}^2(\pi  z).
\end{eqnarray} In writing Eq. \eqref{ijmdbetaa}, we have still used the recurrence relation 
\begin{eqnarray}
    f_{\lambda}(a,b) &=& -\frac{1}{2a^2}\left[ \frac{2\lambda-3}{\lambda -1}f_{\lambda -1}(a,b) \right. \nonumber\\
    &+& \left. \frac{1}{2(\lambda-2)(\lambda-1)}\frac{\partial^2}{\partial b^2}f_{\lambda -2}(a,b) \right],
\end{eqnarray} to handle with $\lambda$ values that do not belong to the allowed range of the function $f_{\lambda}(a,b)$.

\begin{figure}
    \includegraphics[scale=0.5]{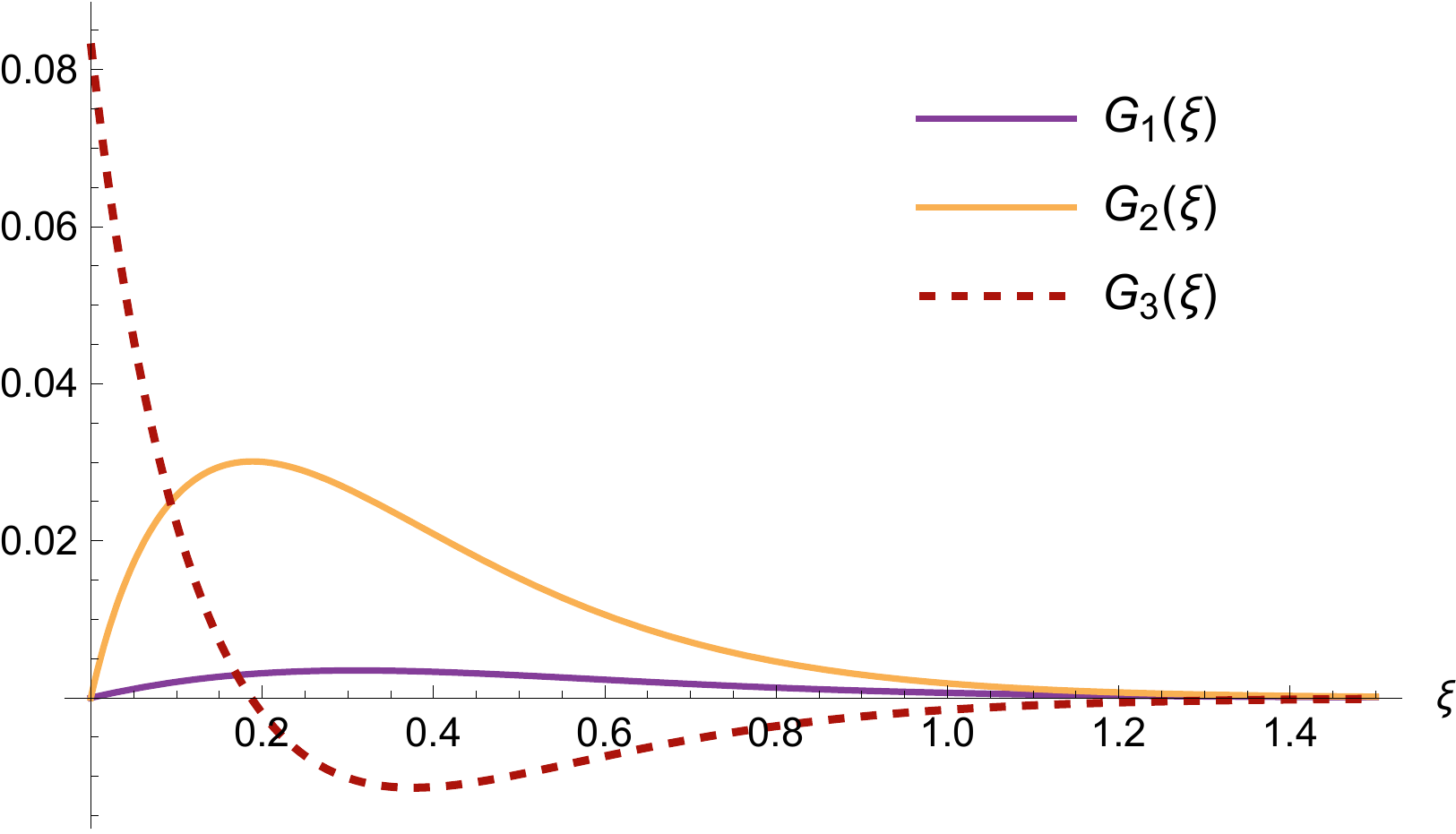}
    \caption{Plot of $G_1$ (purple line), $G_2$ (yellow line) and $G_3$ (dashed red line) as functions of $\xi$.}
    \label{graphfunctiongs}
\end{figure}

Finally, after applying all these results into Eq. \eqref{tddimeni}, we can write the total tensor contribution of the Lorentz violation to the polarization tensor, Eq. \eqref{t1t2t3t4prim}, as
\begin{eqnarray}  
    T_{total}^{\mu \nu}(k) = T_0^{\mu \nu}(k)+T_{\beta}
^{\mu \nu}(k),\end{eqnarray} where
\begin{eqnarray}\label{tzerotempdepend}
   \nonumber T_0^{\mu\nu}(k)&=&-\frac{ie^2}{24\pi^2\epsilon}\left\{k_{\alpha}k_{\beta}(\hat{k}_c)^{\alpha\beta}g^{\mu\nu}+k^2(\hat{k}_c)^{\mu\nu}\right.\\
   &&-\left.k^{\nu}k_{\alpha}(\hat{k}_c)^{\alpha\mu}-k^{\mu}k_{\alpha}(\hat{k}_c)^{\alpha\nu} \right\},
\end{eqnarray} is the temperature-independent part as given by Eq. (\ref{T0}). 
The entire dependence on temperature is contained in the following function 
\begin{eqnarray}\label{tbetatempfinalvec}
    T_{\beta}^{\mu \nu}(k) &=& \frac{i e^2}{\beta^2} \lbrace (\hat{k}_c)^{\mu \nu}H(\xi,k)-(\hat{k}_c)^{\alpha \beta}k_{\alpha}k_{\beta}I^{\mu \nu}(\xi,k) \nonumber\\
    &-& [(\hat{k}_c)^{\mu \alpha}k_{\alpha}k^{\nu}+(\hat{k}_c)^{\nu \alpha}k_{\alpha}k^{\mu}]J(\xi,k) \rbrace,
\end{eqnarray} with
\begin{eqnarray}
    H(\xi,k) &=& H_1(\xi,k)+H_2(\xi,k)+H_3(\xi,k),\\
    I^{\mu \nu}(\xi,k) &=& I^{\mu \nu}_1(\xi,k)+I_2^{\mu \nu}(\xi,k),\\
    J(\xi,k) &=& J_1(\xi,k)+J_2(\xi,k),
\end{eqnarray} and 
\begin{eqnarray}
    \!\!\!\!\!\!\!\!\!\!\!\!\!H_1(\xi,k) &=& \frac{\pi^2}{3}\int_0^1dx\, G_1(\xi)\, (1-x)\nonumber\\
    &\times& \frac{m^4-x(x+2)m^2k^2+x(10x-3)(x-1)^2k^4}{[m^2+k^2x(x-1)]^2},\\
    \!\!\!\!\!\!\!\!\!\!\!\!\!H_2(\xi,k) &=& 2\int_0^1dx\, G_2(\xi)\, (1-x)\,  \frac{m^2+k^2[x(3-2x)-1]}{m^2+k^2x(x-1)},\\
     \!\!\!\!\!\!\!\!\!\!\!\!\!H_3(\xi,k) &=& \frac{4}{3} \int_0^1dx\, G_3(\xi)\, (1-x),\\
   \!\!\!\!\!\!\!\!\!\!\!\!\!I_1^{\mu \nu}(\xi,k) &=&  \pi^2 \int_0^1dx\, G_1(\xi)\, x(x-1)\nonumber\\
   &\times& \frac{k^{\mu}k^{\nu}x(1-2x)^2+g^{\mu \nu}[m^2-2x(x-1)^2k^2]}{[m^2+k^2x(x-1)]^2},\\
   \!\!\!\!\!\!\!\!\!\!\!\!\!I_2^{\mu \nu}(\xi,k) &=& 2g^{\mu \nu}\int_0^1
dx\, G_2(\xi)\, \frac{x(x-1)^2}{m^2+k^2x(x-1)},\\
\!\!\!\!\!\!\!\!\!\!\!\!\!J_1(\xi,k) &=& \frac{\pi^2}{2}\int_0^1dx\, G_1(\xi)\, \frac{k^2x(x-1)(1-2x)^3}{[m^2+k^2x(x-1)]^2},\\
\!\!\!\!\!\!\!\!\!\!\!\!\!J_2(\xi,k) &=& \int_0^1dx\, G_2(\xi)\, \frac{(x-1)(1-2x)^2}{m^2+k^2x(x-1)}.
\end{eqnarray} The temperature dependent functions $G_1(\xi)$, $G_2(\xi)$ and $G_3(\xi)$, with dependence on $\xi$ depicted in Fig. \ref{graphfunctiongs}, are respectively given by 
\begin{eqnarray}
    G_1(\xi) &=& \int_{|\xi|}^{\infty} dz\, \xi^2 \sqrt{z^2-\xi^2}\coth(\pi z) \, \text{csch}^2(\pi  z),\\
     G_2(\xi) &=& \int_{|\xi|}^{\infty} dz\, \frac{\xi^2}{\sqrt{z^2-\xi^2}}[\coth(\pi z)-1],\\
    G_3(\xi) &=& \int_{|\xi|}^{\infty} dz\, \frac{z^2-2\xi^2}{\sqrt{z^2-\xi^2}}[\coth(\pi z)-1].
\end{eqnarray} Note that the temperature-dependent part does not exhibit divergence in the limit $d \rightarrow 3$, and we have assumed this value into Eq. \eqref{tbetatempfinalvec}. From the graph in Fig. \ref{graphfunctiongs}, it can also be noticed that all $G$ functions go to zero at the zero temperature limit ($\xi \rightarrow \infty$), thus recovering the result in Eq. \eqref{tzerotempdepend}, once $T_\beta^{\mu \nu} \rightarrow 0$ as expected. At the high temperatures limit ($\xi \rightarrow 0$), in turn, the function $G_3(\xi)$  is dominant and our result in Eq. \eqref{tbetatempfinalvec} become a quadratic function on temperature, i.e., 
\begin{eqnarray}
   T_{\beta}^{\mu \nu}(k) \rightarrow \frac{ie^2 }{18}T^2\, (\hat{k}_c)^{\mu \nu}.
\end{eqnarray}
Also, it is important to highlight here that the temperature effect does not affect the tensorial structure of the polarization tensor. Now, let us investigate the contributions emergent from the $CPT$-odd terms. 

\subsection{$\hat{k}_a^{\mu}$ contributions}

Analogously, the diagrams for the CPT-odd contributions to the photon self-energy have the same general structure as the CPT-even contributions, just with a different vertex factor, which can be extracted from the last term in the Lagrangian density (\ref{Lag}), and with the diagram $(e)$, from Fig. \ref{figkc}, missing. Consequently, the polarization tensor taking into account
contributions from the Lorentz-violating vector $\hat{k}_a^{\mu}$ can be written as follows:

\begin{equation}
R^{\mu\nu}(k)=R_1^{\mu\nu}(k)+R_2^{\mu\nu}(k)+R_3^{\mu\nu}(k),\label{R0}
\end{equation} where
\begin{eqnarray}
\label{eqrmn1}
    \!\!\!\!\!\!\!\!\!\!R_1^{\mu \nu}(k) &=& -e^2(\hat{k}_a)^{\mu}\int\frac{d^4p}{(2\pi)^4}\frac{(p^{\nu}+q^{\nu})(p^2-m^2)}{(q^2-m^2)(p^2-m^2)^2},\\
    \label{eqrmn2}
    \!\!\!\!\!\!\!\!\!\! R_2^{\mu \nu}(k) &=& -e^2(\hat{k}_a)^{\nu}\int\frac{d^4p}{(2\pi)^4}\frac{(p^{\mu}+q^{\mu})(p^2-m^2)}{(q^2-m^2)(p^2-m^2)^2},\\
    \label{eqrmn3}
 \!\!\!\!\!\!\!\!\!\! R_3^{\mu \nu}(k) &=& -2e^2g^{\mu \nu}(\hat{k}_a)^{\alpha} \int\frac{d^4p}{(2\pi)^4}\frac{p_{\alpha}(q^2-m^2)}{(q^2-m^2)(p^2-m^2)^2}.
\end{eqnarray} Here is another observation: expressions coming from the diagrams $(a)$ and $(c)$ cancel to each other, and Eqs. \eqref{eqrmn1}-\eqref{eqrmn3} refer to the diagrams $(b)$, $(d)$, and $(f)$, respectively.
Also, following step by step the procedures described in the previous section, we are led to the result
\begin{eqnarray}
    R_{total}^{\mu \nu}(k) =R_0^{\mu \nu}(k)+R_{\beta}^{\mu \nu}(k),
\end{eqnarray} with
\begin{eqnarray}\label{tensorpolreszero}
    R_0^{\mu \nu}(k) = 0,
\end{eqnarray} being the polarization tensor at zero temperature and 
\begin{eqnarray}\label{rbetamunufinal}
   R_{\beta}^{\mu \nu}(k) = \frac{ie^2}{\beta^2}\lbrace g^{\mu \nu}(\hat{k}_a)^{\alpha}k_{\alpha}K(\xi,k) + [(\hat{k}_a)^{\mu}k^{\nu}+(\hat{k}_a)^{\nu}k^{\mu}]\, L(\xi,k) \rbrace ,\nonumber\\
\end{eqnarray} the correction due to the temperature. In this last equation, we have
\begin{eqnarray}
    K(\xi,k) &=& K_1(\xi,k)+K_2(\xi,k),\\
    L(\xi,k) &=& L_1(\xi,k)+K_2(\xi,k),
\end{eqnarray} and
\begin{eqnarray}
    \!\!\!\!\!\!\!\!\!\!\!K_1(\xi,k) &=& \frac{\pi^2}{2}\int_0^1dx\, G_1(\xi)(x-1)^2\, \frac{m^2+k^2x(5x-3)}{[m^2+k^2x(x-1)]^2},\\
    \!\!\!\!\!\!\!\!\!\!\!K_2(\xi,k) &=& \int_0^1dx\, G_2(\xi)(x-1)\, \frac{1-3x}{m^2+k^2x(x-1)},\\
    \!\!\!\!\!\!\!\!\!\!\!L_1(\xi,k) &=& \frac{\pi^2}{2}\int_0^1dx\, G_1(\xi)x(x-1)\, \frac{m^2+k^2(1-5x+5x^2)}{[m^2+k^2x(x-1)]^2}. \nonumber\\
\end{eqnarray}
Such null result in Eq. \eqref{tensorpolreszero} is in fact expected at zero temperature, and it can be interpreted as a consequence of the charge conjugation symmetry, which leads to a generalization of the Furry theorem. In its usual version for spinor QED, the Furry theorem states that any fermion loop with an odd number of external photon legs has a null amplitude. The effect of the charge conjugation operator is not related to the spacetime but to the fields. Such operator act by interchanging particles and antiparticles, so that $\hat{C}\phi(x,t)\hat{C}=\phi^{\dagger}(x,t)$. However, the net effect of $\hat{C}$ in the Feynman rules can be described by considering $\hat{C}p_{\mu}\hat{C}=-p_{\mu}$. Hence, inserting an identity operator identified as $I=\hat{C}\hat{C}$ between each propagator and vertex, it is possible to see the cancellation of any loop with an odd number of external photon legs for the contributions with $(\hat{k}_c)^{\mu\nu}$, and also the cancellation of any loop with an even number of external photon legs for $\hat{k}_a^{\mu}$. The difference in the Furry theorem for the $(\hat{k}_c)^{\mu\nu}$ and $\hat{k}_a^{\mu}$ is due to the fact that while the $(\hat{k}_c)^{\mu\nu}$ preserves charge symmetry conjugation, the $\hat{k}_a^{\mu}$ breaks it down, as we can see in the table presented in \cite{Furtado:2020olp}. At finite temperatures, such charge conjugation symmetry breaks down with the presence of the non-null quantity found in Eq. \eqref{rbetamunufinal}. However, that symmetry seems to be restored in the high-temperature limit since the functions $G_1(\xi)$ and $G_2(\xi)$ go to zero in this limit, which is quite an interesting result. 

In addition, it is important to notice that the coefficient $\hat{k}^{\mu}_a$ plays the role of a generalized chemical potential for charged particles. Indeed  both even and odd CPT-coefficients were considered when studying the effects of Lorentz-violation in systems involving phase transitions leading to Bose-Einstein condensation (BEC) at finite temperature. In Ref. \cite{Casana:2011bv}, for example, the authors found some upper bounds for the CPT-even terms considering a nonrelativistic ideal boson gas, besides to verifying that in the relativistic case the chemical potential has Lorentz-violating contributions. In Ref. \cite{Furtado:2020olp}, it was obtained analytical expressions for the pressure, energy, specific heat and charge density for both $\hat{k}^0_a$ and $\hat{k}^i_a$, as well as a correction for the critical temperature $T_c$ that sets the BEC and in this work it is possible to see clearly the role played by $\hat{k}^{\mu}_a$ as a generalized chemical potential. Thus, our work suggests that new contributions may be present both in a generalized Lorentz violating chemical potential as well as in the critical temperature $T_c$ where BEC happens.


\section{Final Remarks}

In this work, we calculate the one-loop self-energy corrections to the gauge field in scalar electrodynamics modified by Lorentz-violating terms in the framework of the standard model extension (SME). We focus our attention on both $CPT$-even and $CPT$-odd contributions. The kinetic part of the scalar sector contains a $CPT$-even symmetric Lorentz-breaking tensor. The interaction terms include a vector contracted with the usual covariant derivative in a gauge-invariant mode. 

For both $CPT$-even and $CPT$-odd cases, the one-loop radiative corrections were computed using dimensional regularization to deal with the divergences present in the zero-temperature regime. The separation of the zero and finite temperature contributions was achieved by the imaginary time formalism and the use of the Larry Ford recurrence relation to performing the sum over the Matsubara frequencies.

For the $CPT$-even contribution, the temperature-dependent part does not exhibit divergence in the limit $d \rightarrow 3$. From the graph in Fig. \ref{graphfunctiongs}, we may notice that all $G$ functions go to zero at the zero temperature limit ($\xi \rightarrow \infty$), thus recovering the usual result at zero temperature case. Moreover, at high temperatures limit ($\xi \rightarrow 0$), the function $G_3(\xi)$  is dominant, and the polarization tensor becomes a quadratic function on temperature.

For the $CPT$-odd case, we have obtained a vanishing result in the zero temperature case, which is a direct consequence of the Furry theorem. We have also verified a breakdown of the Furry theorem at the finite temperature regime. However, the Furry theorem seems to be restored when $T\rightarrow\infty$. In both $CPT$-even and $CPT$-odd cases, the thermal effects do not affect the tensorial structure. The investigation of the analytical behavior of the induced terms at the origin of the momenta space is an important feature that will be explored in future work.  

\acknowledgments

\hspace{0.5cm} The authors thank the Funda\c{c}\~{a}o Cearense de Apoio ao Desenvolvimento Cient\'{i}fico e Tecnol\'{o}gico (FUNCAP), the Coordena\c{c}\~{a}o de Aperfei\c{c}oamento de Pessoal de N\'{i}vel Superior (CAPES), and the Conselho Nacional de Desenvolvimento Cient\'{i}fico e Tecnol\'{o}gico (CNPq), Grant no. 200879/2022-7 (RVM). R. V. Maluf acknowledges the Departamento de F\'{i}sica Te\`{o}rica de la  Universitat de Val\`{e}ncia for the kind hospitality.

\end{document}